# Modular multi-axis stepper motor driver with remote control for use in microscopy

Mathias S. Fischer[1], David Grass[2], Martin C. Fischer[3,*]

1. Department of Mechanical Engineering, North Carolina State University, Raleigh, NC 27695, USA
2. Department of Chemistry, Duke University, Durham, NC 27708, USA
3. Departments of Chemistry and Physics, Duke University, Durham, NC 27708, USA
* Corresponding author: Martin.Fischer@duke.edu

## Abstract

Mechanical translation of samples along several axes is often required in microscopy, and automated positioning requires motorizing the translation stages. Stepper motors are commonly employed but require specialized driver electronics for reliable operation. Here we describe a low-cost, open-source controller design that drives several stepper motors and implements important safety features, such as monitoring of mechanical limit switches, stall detection, and protective software limits. If rotational encoders in the motors or linear encoders on the stages are available, the controller can monitor the stage position and correct it in a feedback loop. The stages can be controlled with serial commands via USB, or optionally from a remote control box that incorporates a joystick for real-time multi-axis speed control, rotary dials for fine axis positioning, and a display to indicate the stage positions. To ease adoption, we provide driver libraries in Python and C, a driver for the commonly used microscopy control software µManager, and an example graphical user interface for calibration and testing.

# (1) Overview

### Metadata Overview

Main design files: https://github.com/MCFLab/PicoStageDriver
Target group: Students and staff in science labs and core facilities, scientists and trained engineers.
Skills required: Hardware assembly – advanced, software installation – easy.
Replication: Prototype has been tested on two microscopes but has not yet undergone replication.

# Introduction

The need for reliable mechanical positioning of objects is ubiquitous, for example in positioning the nozzle over the bed in 3D printing, sheets of paper in a printer, the read heads in hard disk drives, or in our case, positioning samples under the objective of a high-resolution optical microscope. In microscopy, positioning is most often achieved by coupling rotational motors to the lead screw in a linear motion stage. The simplest motor to use is a DC brush motor, where the rotational speed depends on the applied voltage. While speed control is appropriate for some applications, moving a DC brush motor to a specific location requires a position reference, either in the motor itself or on the device that is being moved. An alternative motor type, the stepper motor, inherently rotates in discrete steps (typically a few hundred steps per revolution) when a current sequence is applied to the motor coils. The discreteness of the steps makes positioning easier but precludes a uniform velocity. Finer positioning resolution and smoother operation can be obtained by diving the inherent steps into smaller sub-steps (microstepping).

Many options are available to create the current sequences required to move stepper motors. Commercial motor drivers by stage manufacturers are often optimized for (but usually specific to) the respective motion stages, are available with several control options, such as serial, ethernet, or joystick control [1-3], but tend to be expensive (up to several thousands of dollars). Our controller development originated from the need to replace an ageing and obsolete commercial motor driver system for four motion stages. Hence, we decided to develop a lower-cost alternative and make it accessible to the community. As a starting point we could choose from a variety of driver options, ranging from discrete dual full-bridge driver chips [4] to dedicated stepper motor driver ICs [5, 6]. Such drivers are implemented in several popular add-on boards (shields) to microcontrollers, e.g., [7-10]. Since most of these drivers have limited functionality, e.g., a lack of a computer interface, remote control option, or encoder feedback, we use a microcontroller to provide such functionality. Many do-it-yourself (DIY) designs [11-13] and lab-based custom designs [14-16] of various sophistication and/completeness are available, but none provided the comprehensive set of features and the open designs we desired.

Here we describe our open-source design of a modular, multi-channel stepper motor driver that implements the option for encoder feedback, computer control, optional remote control via joystick and rotary dials, a position display, and safety features like limit switch monitoring and stall detection. The materials cost is about US $150 for a single-axis motor controller, about $360 (or $90 per axis) for a four-axis controller, and about $35 for the optional remote controller.

# Overall Implementation and design

We designed our stage driver to be able to control multiple axes; the current implementation accommodates four axes, limited by the chosen enclosure, but can be easily extended for more axes. In our design, the low-level control of each motor (and optional motor encoder) is performed by an Analog Devices Trinamic TMC5240 driver board (TMC for short). A microcontroller communicates with all TMCs and serves as the interface between a control computer and the TMC driver boards. The TMCs also monitor limit switches on the stages, which can be used for travel range protection and for referencing the stages (homing). We also implemented a "remote" control, which provides control of the stages without the need for communication with a control computer. The remote controller is in a separate, compact box and contains another microcontroller that reads a joystick for X-Y stage velocity control, one rotary dial per axis for fine motor stepping, a sensitivity adjustment for these controls, and a small display to indicate the axes positions. The TMC boards inherently operate in open-loop mode, i.e., they command motor moves but do not actively monitor the motion. If encoders are present, we implement closed-loop operation in the microcontroller. Finally, to ease adoption, we provide software libraries in C and Python, a device driver library for the commonly used µManager software [17], and corresponding example GUIs that can be used to test operation and calibrate the stages. All design files and software are openly accessible in our online repository.

## Choice of stepper driver

Many options are available to drive a stepper motor. A fairly low-level approach is to use a dual H-bridge driver, like the popular L298N [4] or dedicated stepper driver ICs like the Toshiba TC78H670FTG [5] or Allegro A4988 [6]. While ready-made boards incorporating these ICs are available at low cost (some below $10), they are limited in their functionality, such as the lack of trajectory control and encoder/limit switch monitoring. Some more advanced devices incorporate such functionality, like the CL42T from OMC StepperOnline [18] (~US $40) or the TMC line from Analog Devices Trinamic [19] – we chose the latter for its flexibility in configuration and the availability of board schematics. We utilize the basic functionality of the TMC board, such as position/velocity control, monitoring of the encoder and limit switches, basic stall protection, but do not implement the advanced trajectory, low-noise, or low-energy modes of the IC, since those strongly depend on the attached motor and stage and require careful tuning.

## Motor controller design

The required functionality not offered by the TMC board, such as closed-loop or multi-axis operation, was implemented in the microcontroller of the motor controller. This microcontroller needs to be fast enough to handle communication with the control computer, multiple TMC boards, and the remote controller. The Raspberry Pi Pico 2 microcontroller (Pico for short) has more than enough resources for this task, is widely available at low cost ($5), and is easy to program. Figure 1 shows a simplified schematic of the motor controller.

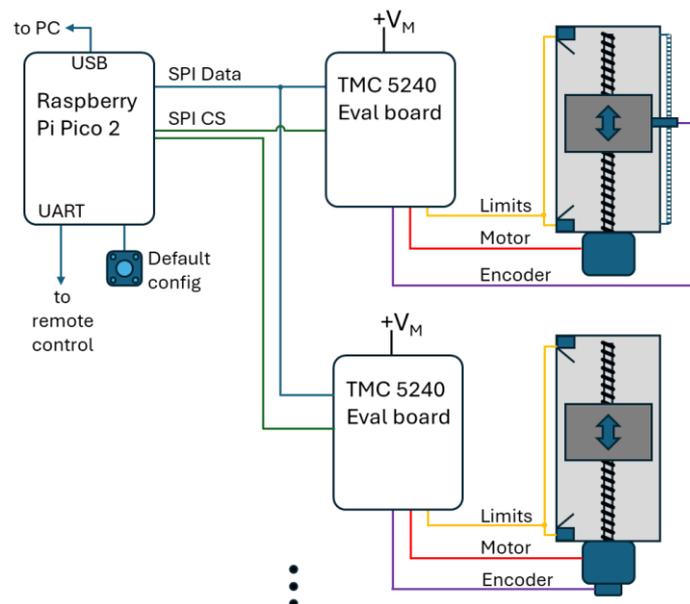

*Figure 1: Simplified schematic of the motor controller. Only two axes are shown, one with a linear encoder on the motion stage, the other with a rotational encoder on the stepper motor.*

Four TMC controller boards are mounted in an aluminum enclosure, all powered by an external power supply (in our case, we utilized a spare power supply from an old laptop computer). Each TMC board is connected to a stepper motor and, optionally, an incremental encoder and limit switches via sub-D connectors at the back of the case (see Figure 2). The encoder and limit switch connections are optional, and the driver can operate without it.

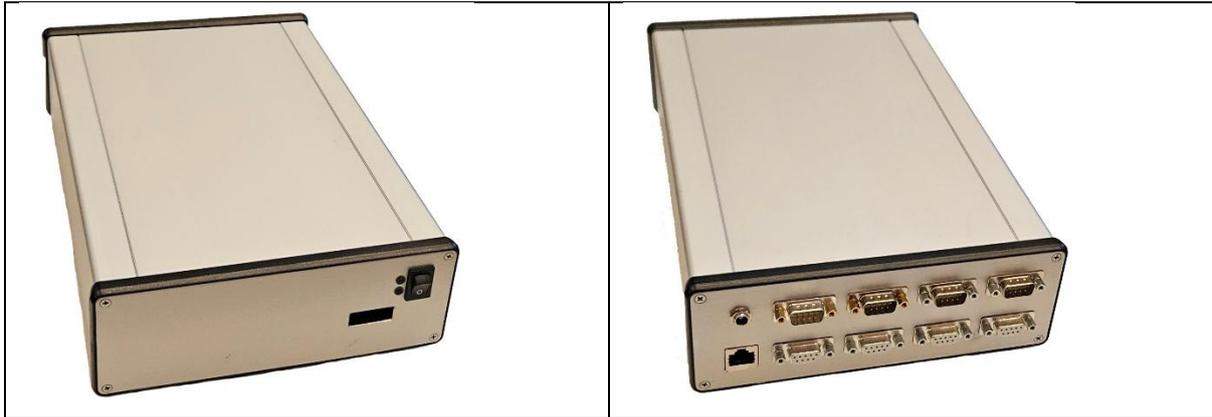

*Figure 2: Picture of the front (left) and back (right) of the motor controller. On the front is a cutout for the USB connector, monitor LEDs for the motor and Pico power, and the switch for the motor power. In the back, there are four sets of 9-pin sub-D connectors, one set for the motors, another (of the opposite gender) for the encoders and limit switches. On the back there is also the power supply connector and the connector for the remote control (RJ45 jack for a standard ethernet cable).*

The Pico communicates with the TMCs using the Serial Peripheral Interface (SPI). The SPI data lines are shared between all TMCs, while each has its own chip select (CS) line. Since the TMCs respond with status bytes to commands sent over SPI, the microcontroller can only communicate with one TMC at a time. Because of the high baud rate used (115200 bd), this is not a practical limitation for common motion stages. The Pico also supplies the TMC boards with 5V and 3.3V for operation of the IC, the encoder, and switches. The encoder can be configured to operate with either 3.3V or 5V.

The Pico handles communication with the computer via serial connection over a USB port. In addition, the Pico communicates with the remote control, if present, over two digital lines via the UART protocol. However, to avoid conflicting positioning commands, only one mode of control (computer or remote) is enabled at any time for each axis.

## Remote controller design

In the remote controller, we included a joystick for two of the stages (in microscopy it is common to have 2-dimensional, or X-Y, stage control), one rotary dial encoder for each of the axis, and a small display to indicate the position of the stages (see Fig. 3). When activated, the joystick sends velocity commands to the X and Y stages, while the rotary dial encoders step the position of the corresponding stage. The maximum velocity in case of joystick control and the maximum step size for dial control are configurable parameters. In addition, we included a one-turn dial (potentiometer) to adjust these maximum values (from zero to their maximum). The joystick and encoders also have a button functionality, which is used to gain remote control for the corresponding axis if computer control is enabled. Switching between joystick and encoder control is also done with a button press. Return to computer control requires a request command from the computer.

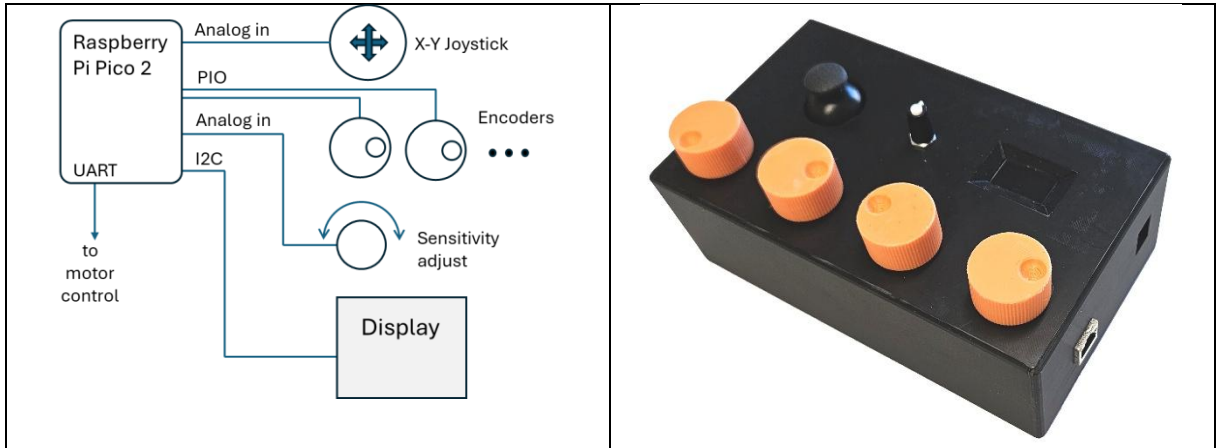

*Figure 3: Schematic (left) and picture (right) of the remote controller. The remote includes a joystick, four rotary dial encoders, a potentiometer for sensitivity adjustment, and a display to indicate stage positions.*

## Motion control modes

The number of steps per revolution in a stepper motor is limited by the mechanical arrangement of the coils (resulting in typically a few hundred steps per revolution). Microstepping is a technique to hold the rotor at (and move the rotor to) positions between the discrete steps by applying variable current levels to the coils, rather than the simple on/off pulses for regular stepping. Hence, the resolution (microsteps per revolution) can be substantially increased. In open-loop operation, a pulse sequence is applied to the motor in order to step (or microstep) the motor by a determined amount. The TMC driver boards operate in this fashion. A desired position trajectory is converted to a pulse sequence, that is then sent to the motor.

However, this open open-loop operation does not actually sense the motion of the motor. It cannot detect if the motor is prevented from rotation due to a mechanical constraint or when the motor is rotated while the coils are not energized. To prevent this uncertainty, stepper motors are often equipped with an encoder that independently measures the angular position of the motor. Alternatively, an encoder can be placed on the motion stage to sense its displacement. Measuring the actual position of the stage also compensates for any nonlinearity in the mapping of steps to position, a common issue with microstepping. We provide an example illustrating such nonlinearity in the general testing section. A common way to compensate for a mismatch in the commanded motor position and the encoder measurement is to calculate the deviation, i.e. the difference between encoder and commanded motor position, and re-command the motor to step by this difference. This process can be repeated until the difference is within a certain tolerance window or a number of tries has been exceeded. We call this mode "open-loop with pull-ins." In this mode, once the desired encoder position has been reached, the controller stops adjusting. Alternatively, in a "closed-loop" mode, these pull-ins can be repeated indefinitely in a loop. In this case, the stage keeps adjusting even after the desired position has been reached. This can be advantageous if thermal drifts or deviations due to mechanical force need to be compensated. In both the "open loop with pull-ins" and the "closed-loop" mode, we apply a simple discrete step correction by the difference only if the difference exceeds a certain tolerance.

# (2) Quality control

## Calibration

The parameters determined during stage calibration are stored in permanent flash memory of the motor controller for automatic reload during boot. Note that parameter adjustment and storage do not require a recompilation and re-flash of the code but can be done via the control computer during normal operation. For initial setup or debugging, pressing the "default load" button during boot of the Pico will load a default set of parameters instead of the set from flash memory. In this section we describe the calibration of the basic motor parameters and defer the description of closed-loop parameters to the general testing section. We only provide a brief description of the most important parameter settings here; more details are provided in the online repository. Also, some of the parameters are directly written to the registers in the TMC5240, so we refer to the TMC datasheet [20] for an in-depth description.

### Motor parameters

The number of microsteps per full step should be determined first, since many of the other parameters depend on it. The TMC can be set from "no microstepping" to 256 microsteps (MS) per full step (FS). Setting this ratio will affect the resolution, i.e. the number of steps per linear stage displacement, in our case steps/μm. A low MS/FS yields a course resolution but generally a more linear mapping of steps to displacement. A high MS/FS yields a fine resolution but exhibits nonlinearities on a fine scale. Also, the resolution is ultimately limited by mechanical constraints like backlash (hysteresis in forward and backward motion) or striction (static friction). Hence, a higher MS/FS count does not guarantee better performance. If an encoder is present, it is a good starting point to set the MS/FS ratio to yield a step size that approximately corresponds to the encoder resolution.

Equally important motor parameters are the max coil currents. The TMC distinguishes "run" current applied during motion of the motor and "hold" current applied after a period if inactivity. These currents are determined by a combination of hardware and register settings in the TMC chip. The data sheet of the stepper motor (or motion stage) should provide an upper limit for the maximum allowable current. The actually required currents will depend on the motor load, the requested speed and acceleration, etc. We empirically choose the run current as the minimum current to still accelerate the motor at the desired rate and the hold current to prevent accidental turning of the motor axis.

The maximum speed and acceleration are usually also specified by the stage manufacturer but need to be converted from linear motion to step rates taking into account the microstep factor (the stage driver units are MS/sec and MS/sec$^2$, respectively).

Optionally, the "StallGuard" feature of the TMC can be activated to disable the motor if excessive load is detected that could lead to a stalling of the motor, thus providing an additional safety mechanism. However, note that this feature will not detect stalls at low velocities and settings need to be empirically determined for each stage. In our experience, the encoder deviation monitoring described below provides a more reliable detection of a stalled motor.

### Switch and homing parameters

The TMC allows for mechanical limit switches at the end of travel of the stages and software-defined ("virtual") limits. For the switches to work, the "direction" of the stages needs to be consistent with switch designation ("left" denoting negative and "right" positive steps) and needs the correct polarity setting (normally open vs. normally closed). The parameters governing the homing procedure (velocity, direction, stop behavior, etc.) should be set and carefully tested at low speeds to ensure correct operation.

### Encoder parameters

If present, an encoder can provide position feedback to ensure the motor reaches the desired position. Having a linear encoder on the stage, as opposed to on the motor, can eliminate stage-related positioning uncertainty, such as backlash of the spindle. To compare the step count of motor and encoder, their relative displacement must be matched. While the step size of the motor can be adjusted by choosing the number of microsteps, the "encoder constant" fine-tunes the ratio of encoder to motor steps. This ratio is best calculated from the motor, stage, and encoder properties, or can be experimentally determined. Once the encoder constant is set, a maximum encoder deviation window can be configured to disable the motor if the difference between motor and encoder step count is exceeded, termed a "following error". A correct encoder constant is also essential for closed-loop operation.

### Remote parameters

The parameters governing the operation of the remote control are the speed at maximum joystick tilt, the step size of one encoder count, and the relative direction of motion for each mode. The maximum speed and step size occur when the potentiometer, or "sensitivity adjust," is at the maximum clockwise position. Because the joystick readings are analog, they are averaged by a running window, and the center position has a definable dead-zone to avoid slow, unintentional stage drift. The zero-position of the joystick is automatically calibrated at bootup and whenever joystick settings are changed.

## General testing

Here we will describe the general testing procedure and operation of the stage driver and compare the open- and closed-loop control modes. As a test system we utilized a bright field microscope (Olympus BX51) that is equipped with an X-Y motor stage (Prior H101A) and a general-purpose stepper motor attached to the Z-focus drive. The X-axis of the motor stage has a 200 steps/rev stepper motor on a 2 mm pitch spindle and limit switches, but no encoder. The Y-axis has the same mechanics but includes an incremental linear encoder (Numerik Jena LIE5 1P N2KV, 0.1 µm scale, 10 encoder counts/µm) mounted on the stage. The Z-axis (objective stage) has a 200 steps/rev stepper motor with a gear system that creates a linear displacement of 100 µm per motor revolution. An external incremental linear encoder (Heidenhain 331666-51, 0.2 µm scale, 20 encoder counts/µm) is mounted on the focus stage. The relevant parameters are summarized in Table 1.

*Table 1: Summary of the stage parameters on our test microscope*

| stage | stepper res [FS/rev] | microstepping [MS/FS] | linear step size [MS/µm] | encoder step size [enc steps/µm] | encoder const [MS/enc step] |
|---|---|---|---|---|---|
| X | 200 | 128 | 12.8 | - | - |
| Y | 200 | 128 | 12.8 | 10 | 1.28 |
| Z (focus) | 200 | 16 | 32 | 20 | 1.6 |

Initially, the microstepping, current, and StallGuard parameters were determined to achieve reliable movement without stalling at the set velocity (here 5 mm/s for X and Y, and 0.5 mm/s for Z). The motor and remote directions of X, Y, and Z were set to match our lab convention, and the switch direction was set accordingly. Virtual limits can be set to ensure safe travel of the stages and prevent damage to the sample, stage, or objective. Note, however, that after homing of a stage, reset of the reference position, or power loss, the virtual limits might need to be reset.

To test the linearity of the stage motion, we moved the motor in equal steps uni-directionally, while measuring displacement with the encoder. Fig. 4 shows the data for the Y-axis (the X-axis does not have an encoder and data for the Z-axis is provided in the online documents).

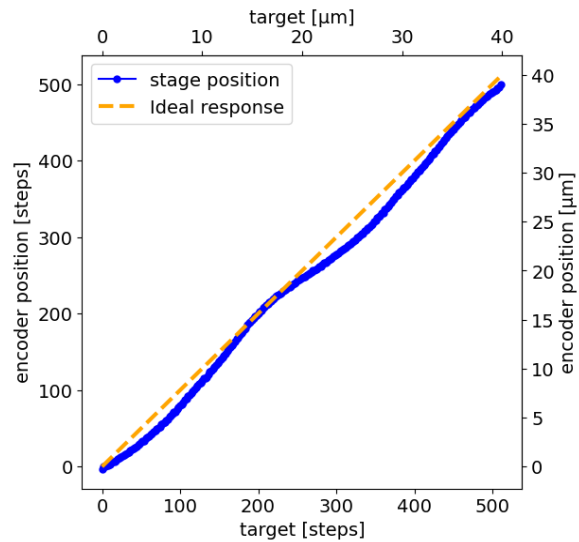

*Figure 4. Increasing unidirectional motion of the Y-stage after an initial movement in the positive direction with a step size of 2 microsteps. The bottom axis is the commanded position of the motor, the left axis the measured position with the encoder. The right and top axis were scaled to μm by the linear step size.*

As can be seen in Fig. 4, the stage position deviates from the expected (ideal) position. In this figure, the total motion extends over 512 microsteps, or 4 full steps. Deviations from linear motion are common in microstepping and even full steps can deviate due to manufacturing tolerances.

Even larger deviations are expected for bidirectional motion, for example due to backlash in the lead screw and other mechanical components. Figure 5 illustrates this for the same stage, where we moved the stage back and forth by the same amount twice, again plotting the actual (encoder) position versus the target motor position.

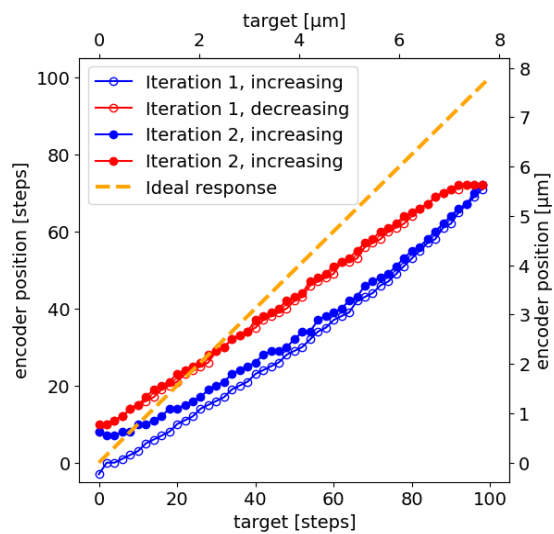

*Figure 5. Bidirectional motion of the Y-stage with a step size of 2 microsteps. After an initial motion in the positive direction, the stage was first moved in the increasing, then decreasing direction for two iterations.*

A hysteresis for forward and backward motion, apparent as non-overlapping increasing and decreasing trajectories, can be clearly seen in Fig. 5. In addition, the increasing trajectories in the first and second iteration differ because of the different starting conditions: in the first iteration the motion was started after an initial motion in the positive direction, in the second iteration it started after the negative motion of the first iteration. The hysteresis, i.e. the deviation of the actual position for the same commanded position in forward and backward direction, exceeds 10 MS for many of the target positions.

Both the hysteresis and the deviation of the motion from linearity can be reduced by employing pull-in moves. For a pull-in, the encoder is read after the motion has stopped, the difference of the encoder position and the desired position is calculated, and the motor is again commanded to a new position that compensates for the latest deviation. The pull-in procedure is stopped when the encoder position is within a given tolerance window, i.e. when the absolute value of the deviation is less or equal than the tolerance, or when a maximum number of tries has been exceeded.

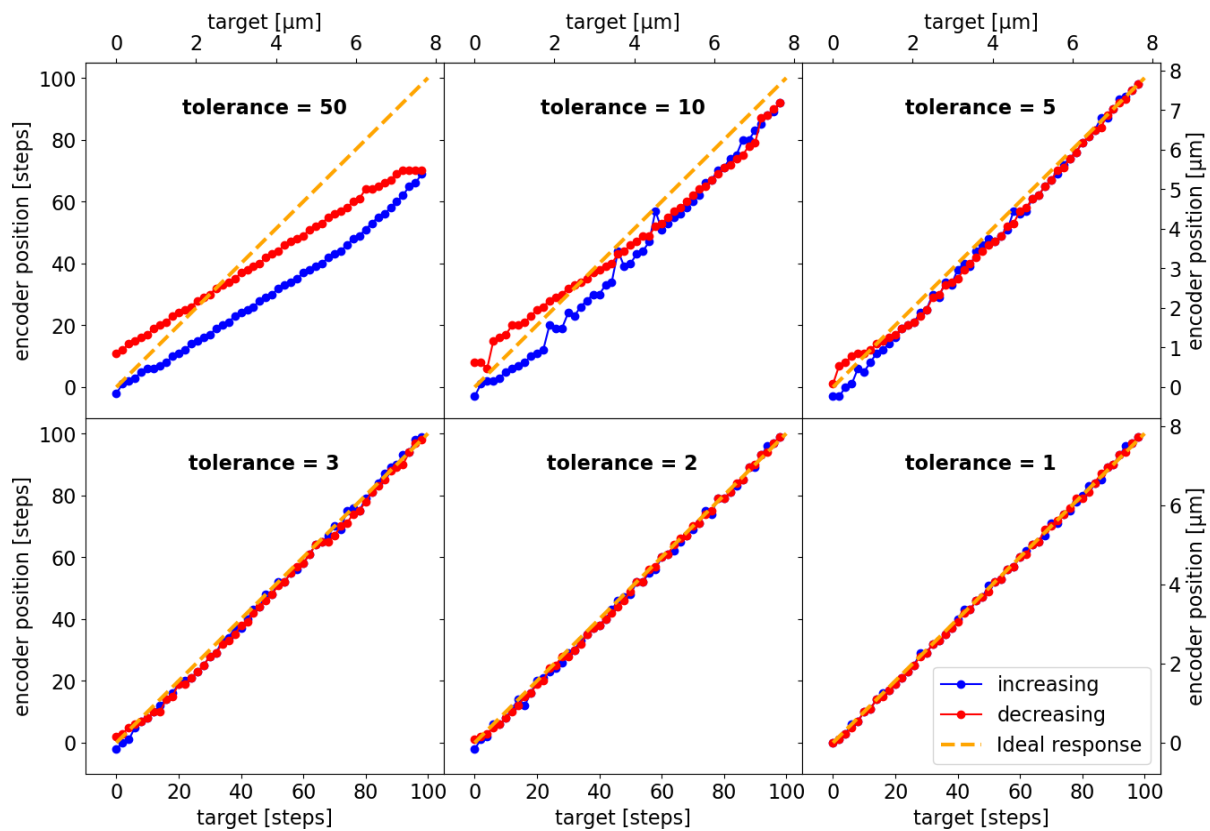

*Figure 6. Bidirectional motion of the Y-stage with a step size of 2 microsteps in the "open-loop with pull-ins" mode. After an initial motion in the positive direction, the stage was first moved in the increasing, then decreasing direction. The chosen tolerance window size (in microsteps) is displayed in the panels.*

The panels in Fig. 6 show bidirectional moves with varying tolerances – a very large tolerance of 50 MS is essentially a simple open loop move. It is apparent that a tighter tolerance reduces the deviation from the requested position. After each "open-loop with pull-ins" move, the stage driver can provide the number of tries to reach the tolerance window. Table 2 summarizes the average deviation and the number of tries for the data in Fig. 6.

Table 2. Magnitude of the deviation between target and measured position, the number of tries to reach the tolerance window, and the total time required to complete the entire run for each panel shown in Fig. 6. Values are given as mean ± standard deviation. Note that the deviation is measured after the pull-ins are complete, hence the deviation can slightly exceed the tolerance due to stage or encoder noise. The last row was acquired with the reset flag enabled (see below).

| tolerance [MS] | \|deviation\| [MS] | max \|deviation\| [MS] | tries | max tries | Time to complete [s] |
|---|---|---|---|---|---|
| 50 | 14.3 ± 9.3 | 30 | 1 ± 0 | 1 | 15 |
| 10 | 6.7 ± 2.8 | 11 | 1.4 ± 0.5 | 3 | 16 |
| 5 | 2.9 ± 1.7 | 5 | 2.7 ± 0.8 | 4 | 18 |
| 3 | 1.7 ± 1.0 | 4 | 3.5 ± 0.9 | 5 | 22 |
| 2 | 1.2 ± 0.8 | 4 | 3.9 ± 1.0 | 5 | 24 |
| 1 | 0.7 ± 0.5 | 2 | 4.8 ± 1.1 | 7 | 33 |
| 1 with reset | 0.6 ± 0.5 | 1 | 1.3 ± 0.8 | 6 | 17 |

From Table 2 we can see that for a tighter tolerance, the deviations decrease, but the number of attempts, and with it the time for motion completion, increases. Note that for the largest tolerance window, there was essentially no wait time since the step size is well within the window. The run time of 15 s therefore is the baseline and is essentially taken up by processing the move commands and checking for errors. Note that no attempts were made to optimize the feedback loop in the "closed-loop" and "open-loop with pull-ins" mode beyond the relay control with tolerance, which proved adequate for our stages.

For remote control with the rotary dial encoder, in which incremental positioning commands are sent to the motor in rapid succession, the large number of tries for small tolerances can create a noticeable response delay. This is especially pronounced if there is an initial mismatch of the TMC motor target and the encoder position. Even though this mismatch is compensated for during the pull-ins, the first (pre-pull-in) move command for a small position change might actually move the stage away from the target position rather than towards it. Hence, we included a flag to optionally reset the TMC motor position to the encoder position after each completed move. This does not affect deviation but reduces the number of required pull-ins. For example, enabling this flag for the tightest tolerance in Table 2 did not noticeably increase deviation, but drastically reduced the number of required tries and time for completion.

We also noticed that our Y-axis stage shows a settling behavior, i.e., after moving the stage to a given encoder position, we sometimes notice a drift away from this position right after the move (over a few seconds). Figure 7 shows this behavior for a set of moves to random positions. The figure shows deviations between the "Open loop with pull-ins" target positions and the encoder positions right after the pull-ins and after a 2 s delay. The closed loop mode (also shown in the figure) avoids this drift since it continuously re-adjusts the stage.

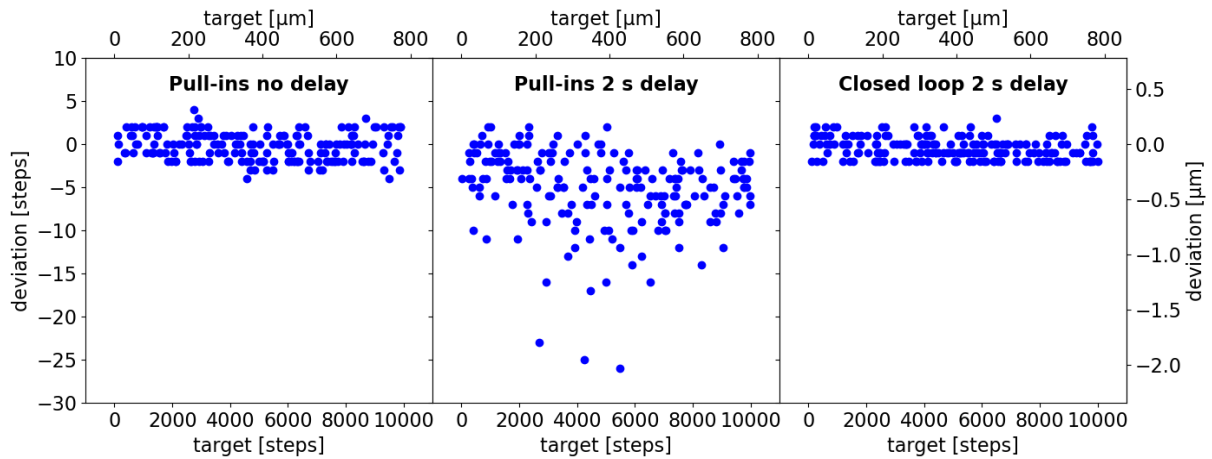

*Figure 7. Deviations between the set and the actual encoder positions after a set of random moves. Shown are measurements in the "Open loop with pull-ins" right after the pull-ins complete ("Pull-ins no delay") and for the same mode 2 s after the moves complete ("Pull-ins 2 s delay"). On the right we show the performance in the closed-loop mode, measured 2 s after issuing the move command ("Closed loop 2 s delay").*

# (3) Application

## Tile acquisition in an optical microscope

To demonstrate the motorized stage motion capability, we performed tiled acquisitions, or "mosaics", of a reference grid sample in a microscope. We imaged a 10 μm square grid target (Thorlabs R1L3S3P) in transmission (see Fig. 8a for a widefield image of the target). We repeatedly imaged the target after moving it in increments of 10 μm in the X- and/or Y-directions in a snake pattern (Fig. 8b). From all the acquired images we extracted image blocks corresponding to a 20 μm x 20 μm area using identical pixel coordinates. We then stacked the blocks side-by-side in the same order as the snake pattern of the acquisition. This arrangement highlights misalignment in the stage position after the movements.

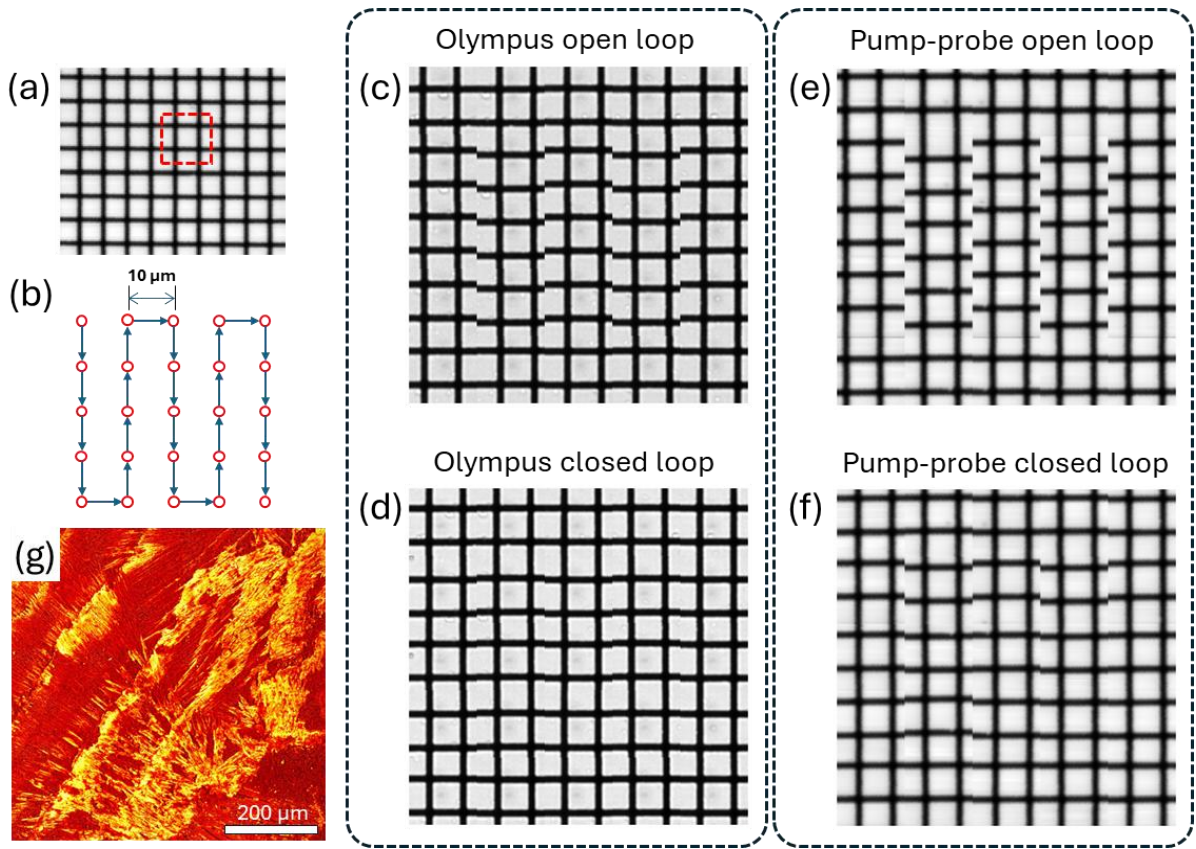

*Figure 8. (a) Widefield image of the grid target. Grid spacing is 10 µm. A block of pixels is extracted from each image in the mosaic acquisition and reassembled as mosaics. (b) Order of tile acquisition. (c) Mosaic assembled from images acquired on our custom Olympus microscope with the stages in open-loop mode. The tiles are blocks corresponding to 20 µm x 20 µm extracted from the images at fixed pixel coordinates. (d) Mosaic acquired on the Olympus, but with the stages in closed-loop mode. (e) Mosaic acquired on our custom pump-probe microscope with the stages in open-loop mode. (d) Mosaic acquired on the P-P microscope, but with the stages in closed-loop mode. (g) Pump-probe mosaic image of $KMnO_4$.*

Figures 8c and 8d were acquired with our custom Olympus microscope described in the General Testing section above. The acquisition in this microscope is orchestrated by µManager and the stages are controlled using our stage driver. In the open loop mode (Fig. 8c) we can see a misalignment between the columns of blocks, apparent as an offset in the Y-direction (the line thickness of the grid is about 2 µm). The misalignment is about 1-2 µm in the central blocks, consistent with the backlash already observed in the general testing section. Because the blocks in the top and bottom rows are always acquired after a move in the same Y-direction, no backlash mismatch is overserved in these rows. The spacing of all but the first columns is uniform because the stage always moves in the same X-direction – the first column can have a backlash depending on the X-position of the stage before the acquisition was started. Figure 8d shows the same acquisition pattern, but with the Y-stage in closed-loop mode. The X-stage is still in open-loop mode since this microscope does not have an incremental position encoder on that axis. The backlash mismatch is greatly reduced due to the position feedback. The remaining mismatch is consistent with the 0.1 µm positional resolution of the incremental encoder.

We also performed a mosaic acquisition with the same grid target on a home-built laser-scanning microscope by scanning the focus of a laser beam across the sample and recording the transmission with a large-area photodiode. The microscope is equipped with an X-Y motor stage (Danaher XYMR-8080), where each axis has a 200 cts/rev stepper motor on a 2 mm pitch spindle and a 2000 cts/rev incremental rotary encoder built into the stage. The Z-stage has a 200 cts/rev stepper motor with a 4000 cts/rev incremental rotary encoder on the Z-focus drive with a linear displacement of 200 µm/rev. We use 16x microstepping for the X- and Y-axis, and 32x for the Z-stage, resulting in a step size of 0.625 µm/MS and 0.03125 µm/MS, respectively. Figure 8e shows the assembled mosaic with the stages in open-loop mode, indicating a substantial misalignment due to backlash of up to ~5 μm, which is less pronounced in closed-loop mode (Fig. 8f).

As a demonstration of full mosaic reconstruction, we acquired a series of transient absorption images of potassium permanganate ($KMnO_4$) with this laser-scanning pump-probe microscope [21]. The $KMnO_4$ was crystallized on a microscope slide and serves as a routine alignment and validation target. We acquired a 6×6 mosaic acquisition with a tile size of 180 μm and reconstructed the mosaic using the "Grid/collection stitching" plugin [22] in ImageJ [23] in linear blending mode (Fig. 8g).

# (4) Build Details

## Availability of materials and methods

The core of our design is the Analog Devices Trinamic TMC stepper control IC, one per stage axis. Several models and form factors (bare chip, breadboard, or evaluation board versions) of these chips are available. For our design we chose a single-channel TMC in the "recommended for new designs" life cycle, that supports the SPI interface, has incremental encoder input, and is offered as an evaluation board. Note that for the TMC chips, the manufacturer provides register mapping in a hardware abstraction layer [24], hence the software can be adapted to a different chip, should this be necessary.

The other central elements of our design are the two microcontrollers, one in the motor controller, the other in the remote control. The resource demands for this project are quite moderate, with several digital and analog IO ports, the most taxing being the simultaneous reading of several rotary dial encoders. The Raspberry Pi Pico 2 is widely available at low cost, is very well supported by the community, and can easily handle these tasks with room to spare (we only use one of the cores). It is, however, by no means the only choice; many low-cost microcontrollers are suitable, and the bulk of our software can easily be ported to another platform.

The remaining components are general-purpose electronic and electrical parts which can easily be substituted with alternate parts should one become unavailable or obsolete. For the motor controller we chose a general-purpose aluminum enclosure for stability; for the remote controller, we designed and 3d-printed a custom plastic enclosure to have more flexibility when mounting the various user IO components, like the joystick, encoders, and display.

## Ease of build / Design decision

For the stage driver, we implemented a modular approach – only the required number of TMC boards need to be included in the build. Our current design supports a maximum of four channels, limited by the chosen enclosure and the maximum current of the power supply, but the Pico in the motor controller has enough spare hardware pins for several (at least four) more. The Pico in the remote controller has enough hardware pins for two more axes, before a re-design is required. Note that the inclusion of the remote control is optional and can be omitted if not required.

A design decision that added cost, but eased construction, was to utilize TMC evaluation boards. As of Oct 2025, the TMCs can be purchased as individual chips (~$8 in single quantity), as breakout board versions (~$20), or as evaluation boards (~$70). Using individual chips, while most cost-effective, requires careful circuit board design, layout, fabrication, and high-density soldering. Few labs have the required resources and for low quantities, outsourcing the production likely offsets the lower chip cost. The use of the breakout board version would alleviate some of the high-density layout and

soldering demands but would still require the design of a custom circuit board to accept the breakout board. For ease of implementation, we chose to use the evaluation board, which comes connectorized, includes protection circuitry for the digital inputs, offers overvoltage protection to dissipate back EMF for larger motors, and does not require chip-scale soldering.

The bulk of the wiring task while building the motor controller is the connectors between the motor controller Pico and the TMC evaluation boards. Even though only 12 of the pins on the TMC connector are used, and some of them are connected locally on the connector itself, 8 wires still connect each TMC board to the Pico. To avoid the need to design and fabricate a custom circuit board, we opted to solder wires to the connector on the TMC side and to use Dupont style connectors and headers on the Pico side. On the Pico side, we used a small protoboard to break out the Pico pins to ease soldering the headers for the connections. All boards in the motor and remote controller are connectorized such that they can be removed individually. Furthermore, the stages are attached to the motor controller via a set of 9-pin sub-D connectors (one for the motor, one for the encoder); we chose the pin layout to be compatible with our existing XY stages, but this can be adapted to each lab's requirements. The remote controller is connected to the motor controller via RJ-45 (8p8c) jack to accept standard ethernet cables to connect the two controllers.

## Operating software and peripherals

To operate the motor controller, the Pico microcontroller needs to be programmed; the same applies to the remote controller, if used. Both are programmed in C++ using the Arduino Pico Framework. The source code can be compiled and flashed onto the Picos either with the Arduino IDE or with Visual Studio Code and PlatformIO. The individual control components are broken into C++ classes, and extensively commented in the source code.

### Motor controller

In the Pico setup phase, component classes are instantiated that handle parameter storage, motor control, serial communication, and remote communication. The setup function also checks the state of the "load default" button; if pressed, the Pico is initialized with default motor parameters, otherwise it loads previously saved parameters from the on-board flash memory. During normal operation, the motor controller Pico continuously loops through four tasks (the time interval for checking each task is adjustable): 1) check for serial commands from the control computer, 2) process any updates to or from the TMC boards, 3) send the current positions to the remote controller display, and 4) check for move requests from the remote controller.

The serial communication task checks for commands coming in via USB from the control computer. Except for the ID query, the commands are ASCII strings starting with the letter "G" (get) for commands that request information from the stage driver, or "S" (set) for commands that send information. "Get" commands return the requested information or an error code. "Set" commands return "ERROR=0" if the command completed successfully or an error code otherwise. Error elaborations are stored for each class, and requesting this verbose error message clears the error flags. Of course, requesting and clearing the error does not remove the source of the errors. The commands are grouped by their function. Parameter commands set or get the motor parameters (current configuration, velocities, limit configuration, etc.) or remote parameters (joystick max velocities, encoder direction, etc.). Motor status commands set or get status values (motor/encoder position, chip temperature, etc.). Motor commands start actions like "travel at a given speed," "move to a specified position," or "start the homing process." Some of these axis-specific commands allow addressing all axes at once, such as enabling or disabling the axes, or checking whether all axes have completed their motion. Pico commands are not axis-specific, such as ID query, error message query, or saving the parameters to flash. For advanced users we also included commands to directly get or set register values on the TMC ICs – please consult with the TMC manual before direct register manipulation. The online repository contains a list of the 60+ commands with explanations and conditions for use (some commands are only allowed when an axis is active and/or enabled, some are disallowed when under remote control, etc.). Note that all commands are non-blocking – for move commands this means that the commend returns after initiating the motion but does not wait for the motion to complete (use the "has position been reached" query for this purpose).

The motor update task coordinates the execution of the move commands, including homing, pull-in, or closed-loop moves. It also prevents conflicting command execution, such as an attempt to move the stage during homing. The task periodically checks the status of the axis, such as the state of the limit switches, the virtual limits, motion completion flags, or following errors. The checks also include error conditions of the TMC IC, such as voltage dropouts, overheating, or a reset after power loss. The "open-loop with pull-ins" and "closed-loop" modes are orchestrated by this task and position updates during such moves are allowed, either via remote computer or remote control. Care was taken to ensure that the motor update task executes fast enough to allow for near-continuous position updates from the rotary dial encoders on the remote controller.

The two remote communication tasks are only performed if remote control is enabled. The first task periodically sends the positions of the active axes to the remote controller for display. The second task checks for incoming commands from the remote controller, such as position or velocity for a specific axis, or the request for remote control of an axis via the button press.

## Remote

In the Pico setup phase, component classes are instantiated that handle the rotary dial encoder, the joystick, the sensitivity adjuster, and the display. For the rotary dial encoders, four state machines in a Programmable IO (PIO) block of the Pico 2 are used to keep track of the dial position. These state machines run independently from the main core and do not require extra resources. For the joystick, we use two analog input pins, one for X and one for Y. The joystick axes provide a voltage from 0 to 3.3 V, which we map to a bi-directional velocity range. Since the joystick does not provide a stable voltage at its center point, we implemented a dead zone of zero velocity around the center position. In addition, single readings are too noisy to reliably control the velocity of the motors, especially around the zero-velocity region. To alleviate this issue, we implemented a moving average of the ADC values, which is running continuously on a repeated timer. After setting up the joystick, we perform an initial joystick calibration to determine the center position and zero-margin. Finally, we set up interrupt service routines to handle the button clicks for both the joystick and the rotary dial encoders.

During normal operation, the remote controller Pico continuously loops through three tasks (the time interval for checking each task is adjustable): 1) check for position and command updates from the motor controller, 2) send position and command updates to the motor controller, and 3) handle the different modes of input and button presses.

The motor controller only issues a few commands for the remote controller to interpret, primarily the joystick and encoder settings (where a change also prompts a joystick recalibration), the motor positions to be displayed, and enabling/disabling of the remote control for each axis. The commands routinely sent from the remote to the motor controller are updates to the velocity if the joystick moves its position (if enabled) or to the position after a turn of the encoder dials (if enabled). The task to handle the input mode simply reacts to the press of a joystick or rotary dial encoder button. After an encoder button press, control is switched and a synchronization of the actual motor position and the dial position is performed. If the motors were initially under computer control, a request is also sent to the motor controller to disallow computer control. Switching back to computer control mode then needs to be initiated from the motor controller.

## Libraries

### Phyton and C

The provided interface library in Python and C handle device initialization, sending/receiving of the get/set commands, and error checking. Table 3 provides a selection of commands from these libraries, intended to show some similarities and differences.

*Table 3. Example commands from the Python and C libraries*

| Python | C |
|---|---|
| `stages = StageDriver("ASRL3::INSTR", logger)` | `status = SD_Init(*hndl, "ASRL3::INSTR");` |
| `stages.set_motor_parameter(motor, "CurrHold", curr)` | `status = SD_SetMotorParameter(hndl, motor, "CurrHold", curr);` |
| `stages.set_motor_command(motor, "Config", None)` | `status = SD_SetMotorCommand(hndl, motor, "Config", 0);` |
| `stages.set_motor_status(motor, "Enabled", 1)` | `SD_SetMotorStatus(hndl, motor, "Enabled", 1);` |
| `stages.set_motor_command(motor, "MoveToPosition", pos)` | `status = SD_SetMotorCommand(hndl, motor, "MoveToPosition", pos);` |
| `stages.set_motor_command(motor, "MoveAtVelocity", vel)` | `status = SD_SetMotorCommand(hndl, motor, "MoveAtVelocity", vel);` |
| `enc_pos = stages.get_motor_status(motor, "EncoderPosition")` | `SD_GetMotorStatus(hndl, motor, "EncoderPosition", &encPos);` |
| `stages.load_parameters_from_file(file_name)` | `status = SD_LoadConfigFromFile(hndl, fileName);` |
| `result = stages.get_error()` | `status = SD_GetErrorMessage(hndl, instrResp);` |

A few notes regarding these examples:

- In the Python library, the initialization function returns a handle to the instance of the driver class, which is then used in successive calls to member functions. In the C library, the handle to the instrument is passed as a pointer.
- In the Python library, the initialization function accepts a logger handle that determines where (and which) error messages are displayed. In the C library, error messages are sent to the console. In both cases, the error messages created in the stage driver should be retrieved after an error is indicated.
- In both cases, the ASCII commands to be sent to the stage driver are translated into a slightly more intuitive format, e.g. "`set_motor_parameter(2, "CurrHold", 10)`" as opposed to "`SMP_CHOL2,10`". A list of corresponding formats for all the commands is provided in the online repository.

### µManager

The device driver for µManager implements the basic functionality to be able to acquire images with stages controlled by the stage driver, such as XY or Z stage motion, enabling and disabling the remote control, resetting the reference positions, and error checking. Automatic device detection and axis assignment by the hardware configuration wizard is implemented. The velocities and accelerations are included as device properties, though by default they are read from the stage driver. We did not include all the motor and remote parameters in the µManager driver, since most of them are only used during calibration. Should access to these be required, we recommend using our GUI.

## Graphical user interface (GUI)

Two simple GUIs, one using the Python library and the other the C library described above, are provided for demonstration and testing purposes. The Python GUI uses the built-in tkinter library, while the C GUI uses the National Instruments LabWindows/CVI development environment (note that CVI is commercial and not freely available). Almost all functionality is included as controls on the GUIs, except for "sensitive" commands, like "saving to flash" or "start homing" to prevent accidental execution. These commands can instead be sent to the stage driver as "direct commands." Figure 7 shows a screen shot of the two GUIs. In the online repository we also provide Jupyter notebooks that were used for acquiring the test data in the "Calibration" section and for data analysis and visualization.

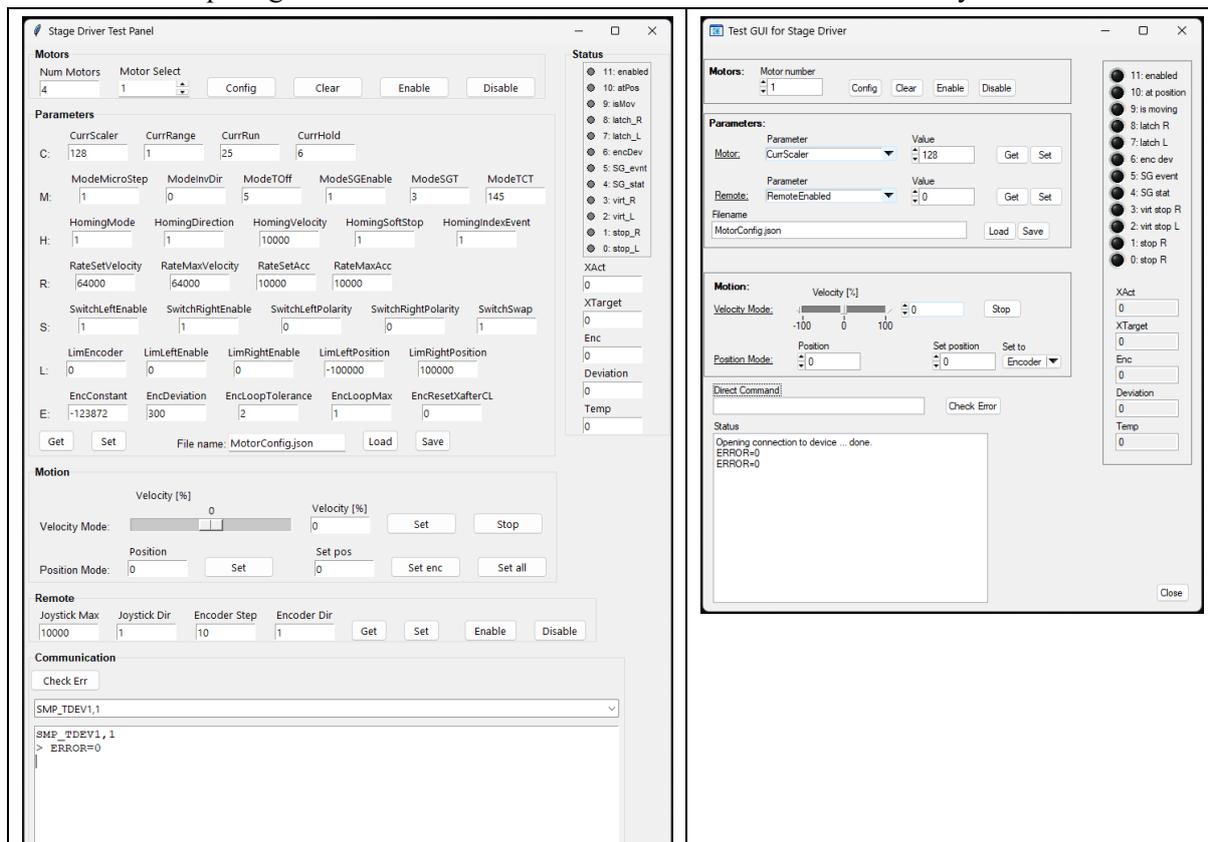

*Figure 9. Screenshot of the Python GUI (left) and C GUI (right).*

## Dependencies

All libraries for building the motor and remote controllers are openly available. Even though the motor driver incorporates commercial TMC stepper controller ICs, the manufacturer provides hardware abstraction source code that is openly available [24]. The remote controller utilizes libraries from Adafruit for the display [25] and the PIO source from Raspberry Pi for the encoder [26]. Both controllers are built either with the Arduino IDE using the Earl Philhower core [27] or using the Gerhard library [28] for Platform IO on Visual Studio Code. The Python and C libraries utilize the Virtual Instrument

Software Architecture (VISA) standard. The Python library utilizes the pyVISA [29] and pyVISA-py [30] libraries, providing an open-source, python-based VISA implementation. The C-library uses the NI-VISA library [31], which is no longer free, but free alternatives are available from several companies, such as Tektronix [32], Keysight [33], or Rohde & Schwarz [34]. The C library also uses the free cJSON library [35] for reading and writing the configuration files.

## Hardware documentation / build instructions / files location:

***Archive for hardware documentation and build files***
***Name:*** https://github.com/MCFLab/PicoStageDriver
***Publisher:*** Martin Fischer
***Date published:*** 11/11/2025

# (5) Discussion

## Conclusions

We have demonstrated a multi-axis stepper driver with remote control. When an encoder is present in the motor or on the motion stage, the position information can be used as feedback for open- and closed-loop operation. We have demonstrated operation on a standard brightfield microscope using the µManager control software and in our custom laser-scanning microscope using our own C-based acquisition software (DukeScan). As of Oct 2025, the total parts cost for our 4-channel controller with remote was approximately $360, which is substantially less than comparable commercial controllers. Our prototype will replace (and upgrade) the now-obsolete driver currently on our microscope (NI PXI-7344 Motion Controller with an NI MID-7604 4-axis driver). We believe that the low cost and freely available software, including a driver for the popular µManager control package, should enable wide dissemination of this device.

## Future Work

A current pain point of the hardware assembly is the need for soldering of a range of connector cables between the remote Pico protoboard and the inputs (joystick, pot, rotary dial encoders), the microcontroller Pico protoboard and the TMC evaluation boards, and the evaluation boards and the motor connectors. Since the motor connectors or their pin assignments are likely to differ between different stages, the latter soldering task is unavoidable. The connection between protoboard and evaluation boards and between protoboard and inputs, however, could be dramatically simplified by designing two interface boards with a socket for the microcontroller and several connectors. In the case of the motor controller we could use ribbon cables, one for each of the TMC boards and for the remote controller Dupont connectors for the prefab cables. Designing these interface boards should be straightforward but would require PCB manufacturing.

The parts cost could be further reduced by utilizing TMC ICs directly, rather than evaluation boards. With some PCB design expertise, a board could be designed that incorporates the microcontroller and several TMC chips on the same PCB, along with the required connectors for the motors and other peripherals. However, because of the high pin density or the no-lead design of the TMC chip (depending on the version used), professional PCB fabrication might be required.

# Paper author contributions (CRediT)

Mathias Fischer: Software, Investigation, Validation, Writing - Reviewing and Editing,

David Grass: Methodology, Writing - Reviewing and Editing,

Martin Fischer: Conceptualization, Methodology, Software, Writing - Original draft preparation, Supervision.

# Acknowledgments

We acknowledge Prof. Warren Warren at Duke University for providing extensive research guidance, mentorship, and lab infrastructure.

# Funding statement

This material is based upon work supported by the National Institute of Health under grant number R21CA280272.

# Competing interests

The authors declare that they have no competing interests.

# (6) Licenses

Article: CC-BY 4.0; Hardware: CERN-OHL-W (weakly reciprocal); Software: GNU General Public License.